\documentclass[11pt,epsfig]{article}  
\usepackage{epsfig}
\usepackage[small,hang]{caption}
\usepackage{cite}  
\usepackage{color}
\newcommand{\beq}{\begin{equation}}
\newcommand{\eeq}{\end{equation}}
\newcommand{\bea}{\begin{eqnarray}}
\newcommand{\eea}{\end{eqnarray}}       
\newcommand{\beqn}{\begin{equation}} 
\newcommand{\eeqn}{\end{equation}}

\newcommand{\al}{\alpha}

\begin{document}
\begin{titlepage}
\noindent
DESY 03-118 \\
\vspace*{2.cm}
\begin{center}
  \begin{LARGE}
    {\bf QCD Corrections to $\gamma\gamma\to ZZ$ \\ at Small
    Scattering Angles}\\
  \end{LARGE}
  \renewcommand{\thefootnote}{\fnsymbol{footnote}}
  \renewcommand{\thefootnote}{\arabic{footnote}}
  \setcounter{footnote}{0}
  \vspace{2.5cm}
  \begin{Large}
    {Grigorios Chachamis$^{(a)}$\footnote{\noindent
     email: grigorios.chachamis@desy.de}}
%    $\quad$
\renewcommand{\thefootnote}{\fnsymbol{footnote}}
\renewcommand{\thefootnote}{\arabic{footnote}}\hspace{-0.5cm}
and {Kriszti\'an Peters$^{(a)}$\footnote{\noindent
        email: krisztian.peters@desy.de}
         $\quad$ }           \\
  \end{Large}
  \vspace{0.3cm}
  \textit{$^{(a)}$II.\ Institut f\"ur Theoretische Physik,
    Universit\"at Hamburg,\\ Luruper Chaussee 149,
    D-22761 Hamburg}  \\
\end{center}
\vspace*{1.5cm}
\begin{abstract}
QCD corrections to the electroweak cross
section of $\gamma\gamma\to ZZ$ at high energies and small scattering angles
have been calculated.
The dominant contributions are due to $t$-channel gluon exchange, i.e.
photons dissociate into quark-antiquark pairs giving
rise to two colour dipoles which interact through gluons.
Corrections resulting from the leading log BFKL amplitude are
of the order of a few percent close to the forward region already at
the 1 TeV energy range and are rising with the scattering energy.
We also considered the helicity non-conserving cases in which the QCD
corrections in comparison to the electroweak part of the amplitude 
strongly grow with energy. The
helicity non-conserving scattering process is of particular interest
since it is sensitive to the Higgs sector.    
\end{abstract}
\end{titlepage}

\noindent
1. The correct extension of the Standard Model (SM) and the
determination of the electroweak symmetry breaking mechanism 
are one of the basic 
questions which have to be answered in the nearest future. 
Experimentally, we expect the first data and insights 
concerning these questions after the run of the Large Hadron Collider
(LHC). Complementary, high precision measurements will come from the
Next Linear Collider (NLC) e.g. TESLA,
opperating in the energy regime up to 1 TeV and providing a very clean
enviroment. In addition one may have the capability of running the NLC
in a $\gamma\gamma$ collision mode via Compton backscattering of laser
photons off the linear collider electrons. Apart from the advantage of the 
higher luminosity, the energy of  
the initial photons can be determined
more accurately than the energy of photons radiated in the $e^+ e^-$ 
collider mode. 
One of the important processes one will consider at the NLC 
is the production of vector bosons such as $\gamma\gamma\to
ZZ$ \cite{Jikia}. 

Concerning the search of physics beyond the Standard Model, 
the fact that the first perturbative
contribution starts at one loop makes the process 
$\gamma\gamma\to ZZ$ sensitive
to a number of investigations. One is the search for the existence of
anomalous triple and quartic vector boson couplings \cite{Anom} or
vector boson Higgs couplings \cite{anomH}. The
natural order of magnitude of these couplings \cite{B} is small, so one
needs to know the SM cross sections with a precision better
than 1\%. 
In order to get a detailed understanding of the spin structure of anomalous 
couplings, it is important to investigate the different helicity states of 
the outgoing Z bosons. 
Because of the absence of the tree level
contribution, this process is sensitive also to particles and new physics
phenomena contributing through radiative corrections \cite{GLPR}. This
leads to a method independent of and 
complementary to the direct production of new particles.  
Again, high precision of the Standard Model cross section is needed.
In addition, the detection of CP violating phases
\cite{CP} or effects due to the exchange of Kaluza-Klein
gravitons in large extra dimension scenarios \cite{ExtraD} have been 
discussed.  

Another motivation to study the process $\gamma\gamma\to ZZ$ is 
its sensitivity to the Higgs sector.  
At high energies the biggest contribution to the cross section comes from the
production of transverse polarised Z bosons in the kinematical limit
of small scattering angles (helicity conserving channel). The dominant 
contribution to the scattering amplitudes is due to  
W loops and grows proportional to the 
scattering energy $s$. Compared to these leading contributions, the 
diagrams containing the Higgs contribution are supressed by $s^2$.
In contrast to this, in the production of longitudinal Z bosons the Higgs 
plays a crucial role.  
In this channel, both the Higgs contribution and the W loop are constant in
$s$ (up to powers of logarithms). 
For large Higgs masses the $s$-channel Higgs exchange in the scattering 
amplitude of $\gamma\gamma\to Z_LZ_L$ violates 
partial-wave unitarity \cite{dobado,boos} and makes this helicity non-conserving case sensitive to the Higgs sector.
Therefore we need to know the SM cross section with a high precision 
in order to disentangle different symmetry breaking scenarios. 

In summary, the process  $\gamma\gamma\to ZZ$ is an important
tool to probe physics beyond the SM. In order to see deviations, 
high precision is
needed, both on the experimental and on the theoretical side. A calculation 
at the lowest available order may not be accurate enough.  
Since the cross section gains its biggest contribution from small scattering
angles, it is natural to ask whether QCD corrections could play a
role in this kinematical regime. At high energies the most dominant 
corrections arise when the vector bosons
fluctuate into quark-antiquark pairs, described by the boson impact
factors, and these dipoles interact
through gluon exchanges. At higher orders in $\alpha_s$ large logarithms in $s$ emerge,
leading to QCD corrections rising with the scattering energy.
 At very high energies these corrections will be large and cannot be neglected.

In this letter we adress the question whether in the energy region of the NLC 
QCD corrections need to be taken into account. For this purpose we 
compute the differential cross section both for the electroweak and the QCD 
parts. 
For the helicity conserving channel the QCD corrections have
been studied recently \cite{petersvacca}. In this letter we
extend our analysis to the helicity non-conserving channel. 
The electroweak part was computed first in \cite{Jikia}, but for our purposes 
we had to repeat the full one loop calculation. 
For the helicity conserving case it turns out that QCD corrections in the region of about 1 TeV are at the percent level and grow moderately with energy. 
For the helicity breaking channel the QCD corrections to the differential cross section, at about 1 TeV, are of the same order, but they grow much faster with increasing energy. 
\\ \\ 

%%%%%%%%%%%%%%%%%%%%%%%%%%%%%%%%%%%%%%%%%%%%%%%%%%%%%%%%%%%%%%%%%%%%%%%%%%%%
\begin{figure}[t]
  \begin{center} \vspace{-0cm}
\epsfig{file=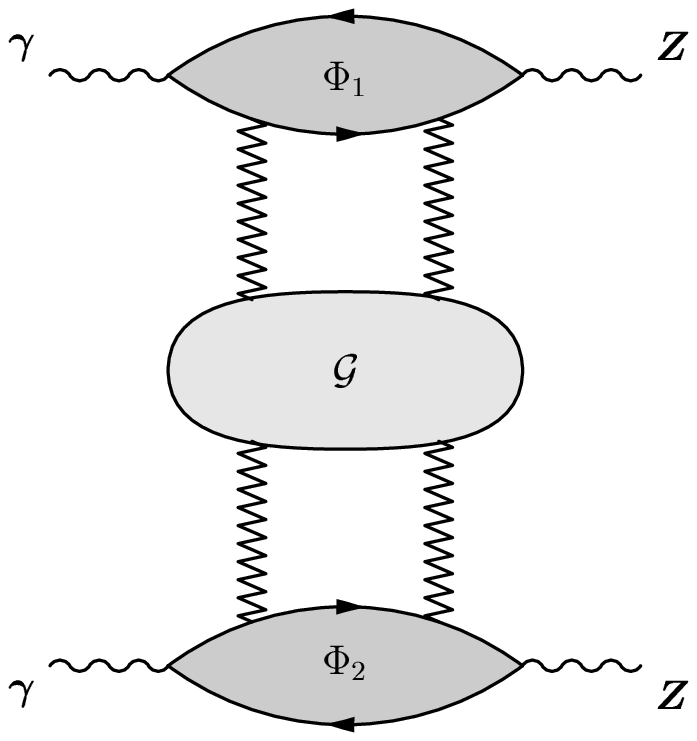,width=5cm,bbllx=192pt,bblly=491pt,bburx=420pt,
bbury=718pt,clip=}
     \caption{The BFKL exchange in the $\gamma\gamma\to ZZ$
        elastic cross section.}
     \label{bfklampl}
\end{center}
 \end{figure}
%%%%%%%%%%%%%%%%%%%%%%%%%%%%%%%%%%%%%%%%%%%%%%%%%%%%%%%%%%%%%%%%%%%%%%%%%%%%% 
\noindent 
2. The differential cross section including the QCD corrections reads:
\begin{equation}
  \label{eq:cross}
  \frac{d\sigma}{dp_T^2}=\frac1{16\pi s^2}|A_{EW}+A_{QCD}|^2 \, ,
\end{equation}
with $p_T$ being the exchanged transverse momenta, 
whereas the corrections to the pure electroweak amplitude are dictated by
the interference term. The calculation was performed in
the Feynman gauge.

The electroweak part of the amplitude was first computed in a full one
loop calculation by Jikia  \cite{Jikia} for Higgs masses over 300 GeV. In
order to compute the above mentioned interference term, the results of
\cite{Jikia} have been reproduced, however we used in this letter an up
to date value of the top quark mass and a Higgs mass of $m_H=115$ GeV.  
We adopted the definition of momenta and polarisation vectors
from \cite{Form}.
The Feynman diagrams were generated with {\em FeynArts} \cite{Feyn} 
and the resulting amplitudes were algebraicaly 
simplified using {\em FormCalc} \cite{Form}. To evaluate the one-loop
integrals the package {\em LoopTools} \cite{Form, Loop} was used.
At small scattering angles the main contribution to the amplitude comes
from the bosonic loop of the helicity conserving channel.
These amplitudes are mainly imaginary and proportional to $s$ in this
kinematical limit, in agreement with the calculation of \cite{petersvacca}
where a high energy approximation was used.
For the helicity-flip channels 
no contributions proportional to $s$ survive, thus these amplitudes
are supressed by one power of $s$ compared to the helicity
conserving cases. If one of the Z bosons has transverse and the other 
longitudinal polarisation, the amplitude is vanishing at the forward
point ($p_T=0$) due to angular momentum conservation.

The QCD part of the amplitude was calculated recently in
\cite{petersvacca}, where the reader will find analytic expressions
for $A_{QCD}$. In the small angle region these emerge when vector
bosons fluctuate into a quark-antiquark pair and these dipoles
interact through gluons. At the lowest order, when  two gluons are
exchanged in the $t$-channel, one obtains a contribution proportional
to the scattering
energy for both, the helicity conserving and helicity flip amplitudes. The
radiation of more gluons enhances the cross section, since this higher
order corrections provide large logarithms in energy, which are rising
with the scattering energy. These contributions cannot be negleted at
large energies. One possibility to take this into consideration is a
resummation described by the LO BFKL equation \cite{BFKL} which gives an
upper bound estimate of these effects. 

The Feynman diagrams for $A_{QCD}$ are illustrated in Fig.1. Due to the
high-energy factorisation one may calculate first the boson
non-forward impact factors $\Phi$ associated to the external particles
and integrate these with the BFKL Green's function ${\cal G}$. The
BFKL Green's function is the result of the resummation of leading
logarithms in energy, coming from diagrams of ladder topology, built
with non-elementary reggeized gluons \cite{BFKL}. The boson impact
factor is the convolution of the two boson wave functions which describes
the probability that a boson fluctuates into a quark-antiquark pair
\cite{peters}. The calculation was done for the kinematical region of
small scattering angles in the high-energy approximation. Thus, one may
neglect terms suppressed by powers of $t/s$, which
simplifies the calculation significantly.  

One important property of the helicity flip impact factors need
particular attention concerning our further calculations: these impact
factors are in
general non-zero, they vanish only for forward scattering, where
$p_T=0$. This comes from a different symmetry behaviour of the
transverse and longitudinal wave functions. Writing these in a
coordinate space formulation  \cite{peters}, the longitudinal wave
function is symmetric under the transformation of the dipole size
vector ${\bf r  \to -r}$, while the transverse one is
antisymmetric. Since the dipole interaction is also symmetric under
this transformation, the convolution of this with a transverse and
longitudinal wave function is antisymmetric, leading to the vanishing
result for $p_T=0$. For non-zero $p_T$, this symmetry
properties are broken, resulting in non-vanishing helicity flip impact
factors. Real and imaginary parts of the helicity flip impact factors
are oscillating with $p_T$ but shifted in a way that the absolute value of the
amplitude gives a smooth function (Fig. 9 of \cite{petersvacca}). 
The helicity flip impact factors are constant in energy, thus the corresponding
amplitudes are proportional to the scattering energy $s$
\cite{petersvacca,peters}. Because the electroweak parts of the
helicity flip amplitude are suppressed by one power of $s$ compared to the QCD
ones, the QCD corrections
will increase rapidly with energy, leading to significant corrections in
the TeV energy regime. \\ \\ 
%%%%%%%%%%%%%%%%%%%%%%%%%%%%%%%%%%%%%%%%%%%%%%%%%

\noindent
3. The order of magnitude of the QCD corrections is determined in a
numerical analysis. Here we consider full circular polarisation of the
incoming photons. For the QCD part of the amplitude, 
due to the huge rapidity separation of the bosons, 
it is only important if the helicity
is conserved or broken in each impact factor. As a result, the
amplitude $++\to TT$ is equal to the amplitude $+-\to TT$ as well as
to the amplitude with unpolarised photons. 
The mass of the Higgs boson was set to $m_H=115$ GeV unless a
different value is stated and the parameters
$\al_W=\al/s_W^2$, $\al =1/128$, $m_Z=91.2$ GeV, $m_t=174.3$ GeV 
and $\alpha_s (M_Z)$
have been used throughout the numerical computations.  
Since in the QCD expressions \cite{petersvacca} the quark masses are 
always accompanied by the Z mass, they can be negleted in the
numerical calculations apart from the top quark mass. The inclusion of the top
quark mass reduces the QCD amplitude by 25\%. 

At high energies in the small angle limit
one expects enhancements due to the appearance of large logarithms,
these have been resummed in the BFKL scheme. The BFKL resummation was
evaluated in the saddle point approximation \cite{petersvacca}.  
The resummed leading log QCD corrections hold an 
uncertainity resulting from the scale which is not fixed at this order
of the calculations. The scale was set to $s_0=M_Z^2$ in the numerical
evaluations. We stress that the resummed leading log QCD corrections
at the lower energies we consider, 
are overshooting what is expected from the true contribution,
nevertheless they provide a first estimate of these corrections.  

In Fig.2(a-d) QCD corrections to the differential cross
sections relative to the pure electorweak contributions are
plotted. These relative corrections are defined as:
\begin{equation}\label{eq:corr}
\Delta=\left(\frac{d\sigma_{QCD+EW}}{dp_T^2} - 
  \frac{d\sigma_{EW}}{dp_T^2}\right)\left/
  \frac{d\sigma_{EW}}{dp_T^2}\right. \, 
\end{equation}
and are presented as functions of $p_T^2/M_Z^2$ for center of mass
energies of 1 Tev and 3 TeV. At these energies  $p_T^2/M_Z^2=4$
corresponds to values of $\cos\theta$ (where $\theta$ is the
scattering angle) of 0.90 and 0.99 respectively.   
Thus, for rising energy the scattering angle will be continously 
smaller for the same $p_T$ values. For a TESLA detector it was
proposed that the tracker system will reach values $\cos\theta <
0.993$, in this
range one has a good measurement possibility \cite{Behnke:2001qq}. 
Moreover, the
decay products of the Z bosons will carry also transverse momenta,
thus these particles can have bigger angles from the beam pipe.

%%%%%%%%%%%%%%%%%%%%%%%%%%%%%%%%%%%%%%%%%%%%%%%%%%%%%%%%%%%%%%%%%%%%%%%%%%%
\begin{figure}\label{fig:corr} 
\hspace{.5cm}  
\epsfig{width=16cm,file=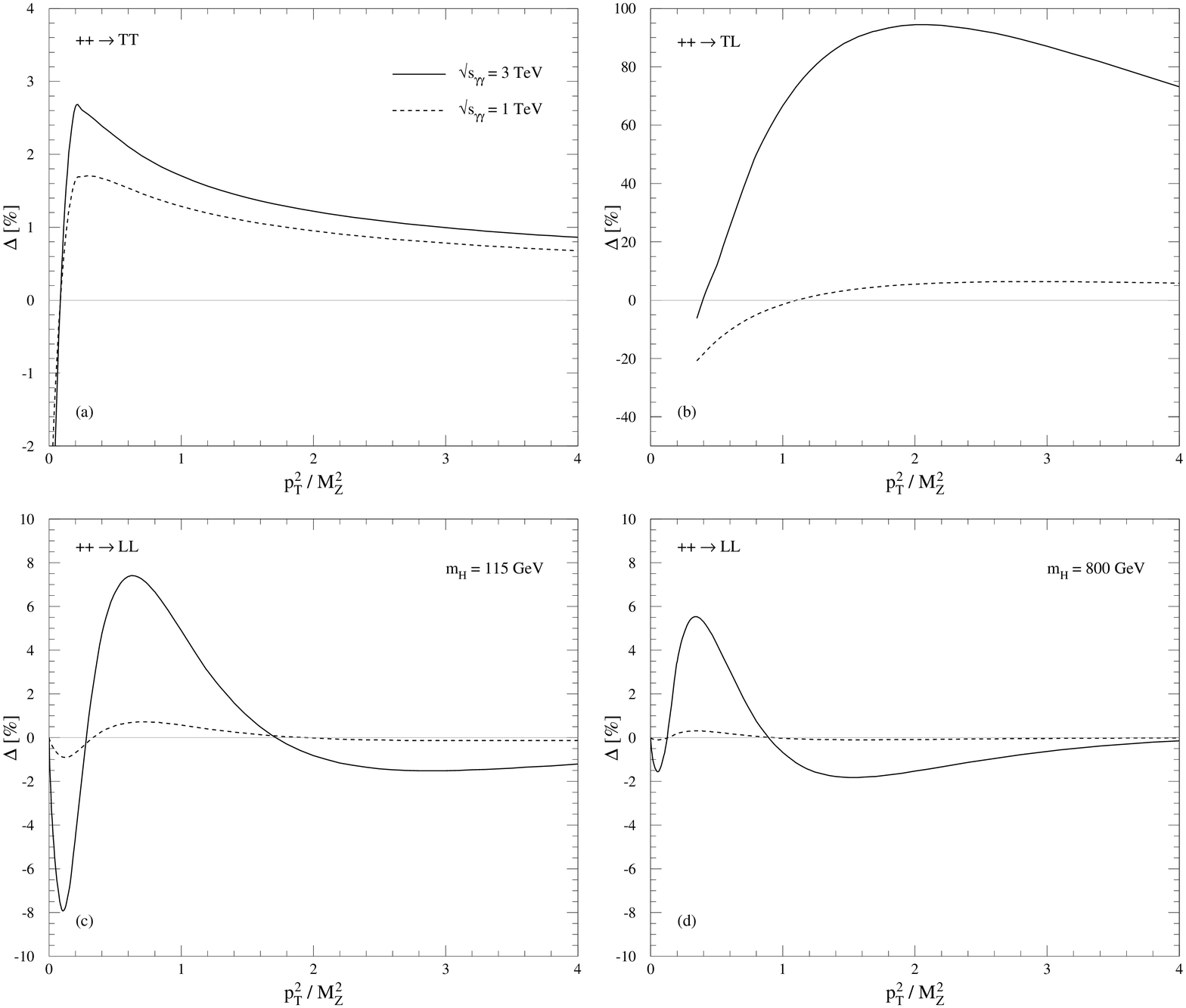}   
\caption{QCD corrections to the differential cross section relative to
  the pure EW contribution for different Z polarisations and center of
  mass energies $\sqrt s$. The relative correction is defined as
  $\Delta= (\frac{d\sigma_{QCD+EW}}{dp_T^2} -
  \frac{d\sigma_{EW}}{dp_T^2})/\frac{d\sigma_{EW}}{dp_T^2}$.}
\end{figure}        
%%%%%%%%%%%%%%%%%%%%%%%%%%%%%%%%%%%%%%%%%%%%%%%%%%%%%%%%%%%%%%%%%%
\begin{figure}[p]\label{fig:intcorr} 
\hspace{1cm}  
\epsfig{width=13.2cm,file=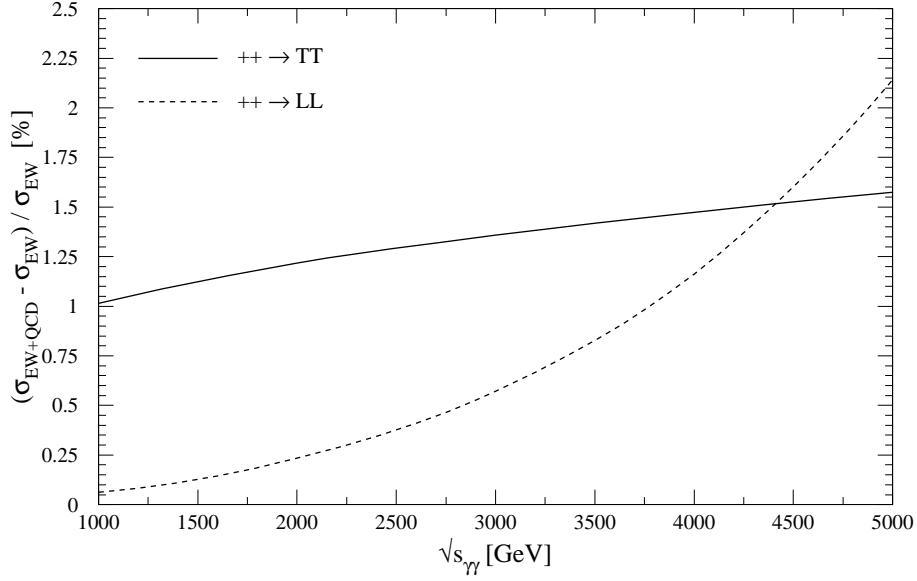}   
\caption{QCD corrections to the integrated cross section relative to
  the pure EW contribution for transverse and longitudinal polarised Z
  bosons. The integration region was choosen as $0 < p_T^2 < 4 M_Z^2$.}
\end{figure}        
%%%%%%%%%%%%%%%%%%%%%%%%%%%%%%%%%%%%%%%%%%%%%%%%%%%%%%%%%%%%%%%%%%
\begin{figure}[p]\label{fig:intcorrtl} 
\hspace{1cm}  
\epsfig{width=13.2cm,file=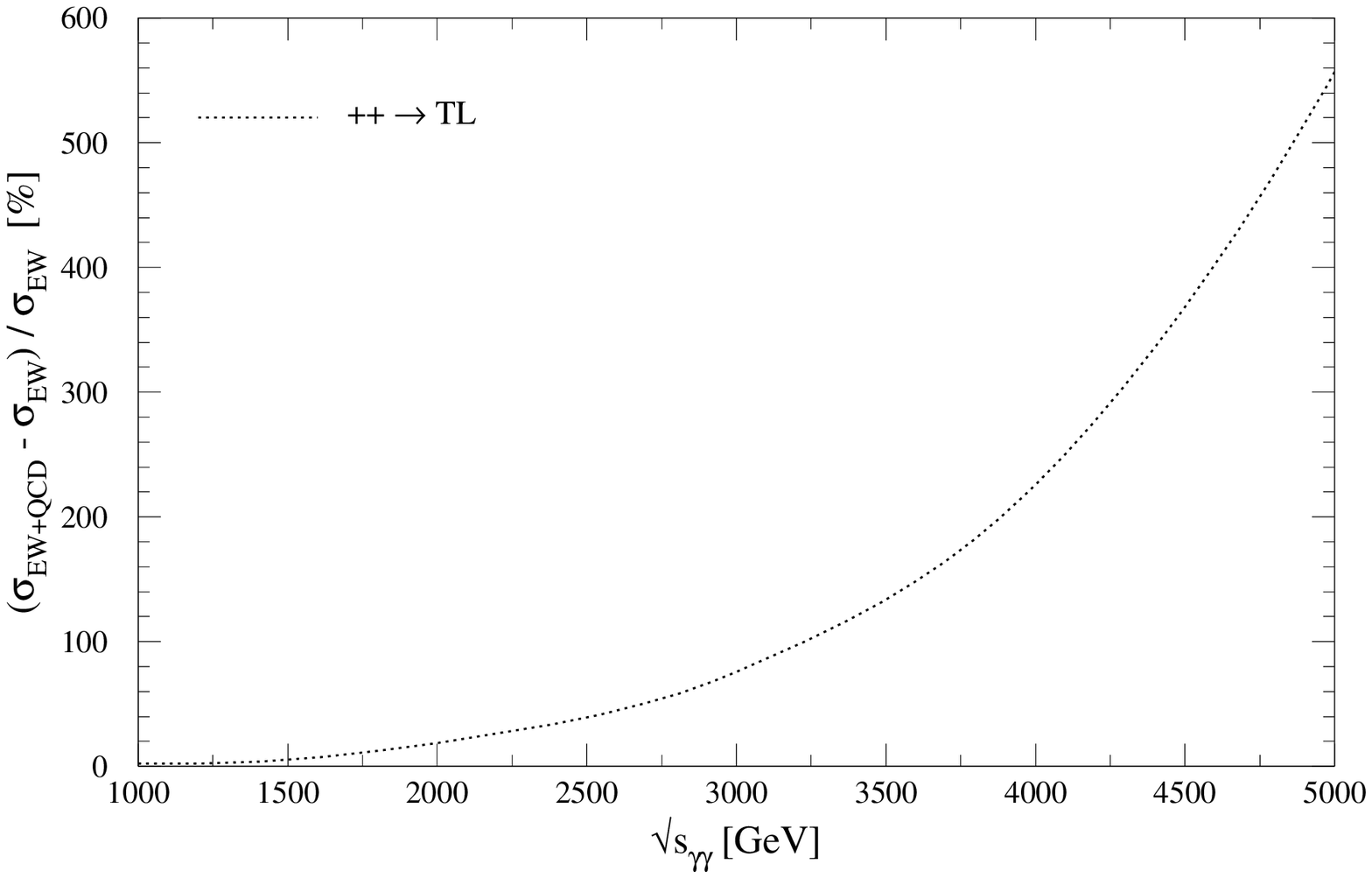}   
\caption{QCD corrections to the integrated cross section relative to
  the pure EW contribution where the Z bosons have different polarisation. 
  The integration region was choosen as $1/2 M_Z^2 < p_T^2 < 4 M_Z^2$.}
\end{figure}        
%%%%%%%%%%%%%%%%%%%%%%%%%%%%%%%%%%%%%%%%%%%%%%%%%%%%%%%%%%%%%%%%%%

Fig.2(a) shows QCD corrections for transverse polarised Z 
bosons. The electroweak amplitude is mainly imaginary, in agreement
with the result of \cite{petersvacca} calculated in the high-energy
approximation. Therefore, from the QCD amplitude it will be also the imaginary
part which mainly accounts for the interference term.   
The relative corrections are of the order of percent level in the helicity
conserving channel. For higher $p_T$ they are approximately one
percent and they are rising up to a few percent while approaching the
forward region, since the QCD amplitude gets more
enhanced compared to the electroweak one.
This is due to the fact that in the transition to forward physics the
perturbative QCD analysis is increasingly affected by the long
distance interactions, since the gluons are massless.
The slight rise of the corrections from 1 to 3 TeV is due to the
leading log BFKL resummation and is dictated by the Pomeron
intercept. Almost at the forward region the imaginary part of the 
QCD amplitude changes sign and becomes negative which is also visible in
the relative corrections. 

Next we discuss the helicity breaking cross sections. As discussed in the previous section, the most
important property of the helicity flip amplitudes is, that the electroweak parts 
are not anymore proportional to $s$ as the helicity conserving parts
were, so they are supressed by one power of $s$ compared to the QCD
part of the amplitude. As a consequence the QCD corrections in
comparison to the electroweak part will
increase rapidly with the scattering energy, they vanish only for $p_T=0$.
This appear in Fig.2(c) where the relative QCD
corrections are plotted for longitudinaly polarised Z bosons. 
The corrections are first rising when $p_T$ is becoming smaller but they
vanish for $p_T=0$. For 1 TeV the corrections are less than 1\%, but
for 3 TeV they are rising up to 8\%. This strong rise is dominating by
the different $s$ dependence between the electroweak and QCD
amplitude. The additional rise coming from the BFKL resummation is
neglible compared to this effect. While the electroweak part of the  
amplitude was mainly imaginary in the helicity conserving 
cases, here real and imaginary parts are of the same order. The
oscillations visible in the plot are coming entirely from the QCD
part of the amplitude, described in the previous section.
For higher Higgs masses the real part of the electroweak amplitude gets
an enhancement from the Higgs pole contribution, thus the form of
the corrections is changing. 
This is illustrated in Fig.2(d) where the calculation was 
done for a Higgs mass of $m_H=800$ GeV. 

In Fig.2(b) relative corrections are plotted where one of
the Z bosons is transverse and the other longitudinally polarised. 
The most striking property of these corrections is their magnitude,
they can be as high as 100\% already at $s=3$ TeV. This is due to the
fact that here the EW part of the amplitude is smaller but the QCD
part is bigger in comparison with the case where both Z bosons are
longitudinally polarised. The enhancement of the QCD part is coming
from the helicity conserving impact factor $\Phi$, which is
separated by a huge rapidity gap from the supressed helicity
flip impact factor. On the other hand, the QCD part is vanishing for
$p_T=0$ since the helicity flip impact factor does. Because of the
dominance of the helicity conserving impact factor, the oscillations
coming from the helicity flip impact factor are not visible any more. 
We observe here again corrections which are rising if $p_T$ gets smaller. 
Both, electroweak and QCD parts of the amplitude have to vanish for the 
case of forward scattering due to angular momentum conservation.   
Since the QCD part has its turning point for bigger $p_T$, approaching the 
forward point the corrections are decreasing. We did not plotted 
corrections up to $p_T=0$ since eq.(\ref{eq:corr}) lacks of definition
at the forward point. Again, here the corrections are rising strongly
with $s$, since the electroweak amplitude is supressed by one 
power of $s$ compared to the QCD one. The relative corrections are
approximately one order of magnitude bigger if $s$ grows from 1 to 3 
TeV. The rise resulting from the BFKL resummation is again negligible in this 
context.

Next we display corrections to the integrated cross section. The high 
enegy approximation on which the QCD calculation is based, is for a 
kinematical range where $s\gg -t$. Thus, for the corrections to the 
integrated cross section we integrated only up to $p_T^2=4 M_Z^2$, since 
for high $p_T$ the QCD calculation is loosing its validity. 
The solid line in Fig.3 displays these corrections for the 
helicity conserving case. In this integration range the corrections are 
around one percent and they have a slight rise due to the BFKL 
resummation, since (up to powers of logarithms) electroweak 
and QCD amplitudes have 
the same behaviour in $s$. The dashed line is for corrections with two 
longitudinal Z bosons in the final state. In this helicity breaking part 
the electorweak amplitude is suppressed by one power of $s$ in
comparison to the QCD amplitude, so the 
corrections are rising rapidly with the scattering energy. The smallness 
of the corrections in the integrated cross section is due to the fact that 
the corrections to the differential cross section, Fig.2(c),  
are changing sign with $p_T$ varying, presenting an oscillating behaviour. 

In Fig.4 again corrections to the helicity flip cross 
section are plotted, but here one of the Z bosons is transverse polarised.
Due to the different $s$ behaviour of the amplitudes again a 
strong rise of the corrections with $s$ is present. These are very big, for higher 
energies the QCD part of the amplitude completely dominates this part of 
the cross section. The magnitude of these corrections is mainly due to
the helicity conserving impact factor  $\Phi$ of the QCD amplitude. 
The integration 
was done down to $p_T^2=1/2 M_Z^2$, because the relative corrections lose 
they meaning for $p_T=0$, since both amplitudes are vanishing at the
forward point. \\ \\ 

\noindent
4. In summary, we have computed QCD corrections for $\gamma\gamma\to ZZ$ at high center of mass energies
   in the kinematical region of small scattering angles. 
We have considered the exchange of BFKL gluon ladders which couple
  to the incoming photons via $\gamma\to Z$ impact factors.
   The electroweak part was computed in a
   full one loop calculation, and a complete analysis involving
   all helicity channels has been done.  

In the helicity conserving channel the corrections are at the order of
a few percent for $s=O$(1 TeV) and they show a moderate rise 
with the scattering energy. 
In the helicity flip channels the QCD corrections are at the same
level for $s=O$(1 TeV). However, since in this channel the electroweak 
amplitudes are suppressed by one power of $s$ compared to the QCD
ones, the corrections rise much 
stronger with the scattering energy. Already for $s=O$(3 TeV)
the QCD corrections are significant.

Concerning the QCD corrections we stress, that the leading log BFKL 
contribution contains a noticable
scale dependence.  For a more precise analysis one has to use 
the next-to-leading BFKL Green's function and next-to-leading impact factors.
One has also to look in the full SM two loop contribution, in order to
achieve precision at the percent level. 

On the experimental side the separation of longitudinal and transverse final state Z bosons clearly presents a demanding challenge.  Only a
statistical analysis of the angular distribution of the decay
products of the Z bosons allowes to discriminate between the different
polarisations. However, as discussed in the beginning of this letter,
a carefull measurement of the process $\gamma\gamma\to ZZ$ with all its different helicity configurations is important and should be persued. The results of our study indicate that in the analysis of the measurements QCD corrections cannot be neglected.
\\ \\ 

\noindent 
{\it Acknowledgements.} We wish to thank J. Bartels for many helpful discussions.
The authors are supported by the \textit{Graduiertenkolleg}
"Zuk\"unftige Entwicklungen in der Teilchenphysik".
%%%%%%%%%%%%%%%%%%%%%%%%%%%%%%%%%%%%%%%%%%%%%%%%%%%%%%%%%%%%%%%%%%%%%%%%%%%%

%%%%%%%%%%%%%%%%%%%%%%%%%%%%%%%%%%%%%%%%%%%%%%%%%%%%%%%%%%%%%%%%%%%%%%%%%%%%

\end{document}